\documentclass[aps,prb,reprint,groupedaddress]{revtex4-1}

\usepackage{graphicx}
\usepackage{epstopdf}
\usepackage{dcolumn}
\usepackage{bm}
\usepackage{color}
\usepackage{natbib}
\usepackage{amsmath}
\usepackage{amssymb}

\begin{document}

\title{$J_{eff}=1/2$ Mott Insulating State in Rh and Ir Fluorides}

\author{Turan Birol and Kristjan Haule}
\affiliation{Department of Physics and Astronomy, Rutgers University, Piscataway 08854, NJ, USA}

\begin{abstract}
Discovery of new transition metal compounds with large spin orbit
coupling (SOC) coexisting with strong electron-electron correlation
among the $d$ electrons is essential for understanding the physics that
emerges from the interplay of these two effects.
In this study, we predict a novel class of $J_{eff}=1/2$ Mott
insulators in a family of fluoride compounds that are previously
synthesized, but not characterized extensively. First principles
calculations in the level of all electron Density Functional Theory +
Dynamical Mean Field Theory (DFT+DMFT) indicate that these
compounds have large Mott gaps and some of them 
exhibit unprecedented proximity to the ideal, $SU(2)$ symmetric $J_{eff}=1/2$ limit.
\end{abstract}

\date{\today}

\maketitle

Interest in $5d$ compounds has been blossoming in the recent years in
response to the scientific advances and applications in the areas of
topological insulators, multiferroics, and thermoelectrics. At the
forefront of this activity are the Ir compounds, because of the
interesting interplay between itinerancy, the electronic correlations
and strong spin-orbit coupling (SOC).\cite{witczak-krempa2014}
%
%
This strong coupling between the spin and orbital degrees of freedom
gives rise to various interesting phases, such as the exotic
spin-liquid phase predicted in honeycomb iridates, or the recently
observed Fermi arcs and the spin-orbit induced Mott insulating phase
in the perovskite-related
Ir-oxides.\cite{witczak-krempa2014,wan2011,moon2008,kim2014} In these
latter systems, the SOC splits the six-fold
degenerate Ir t$_{2g}$ states into 4 occupied $J_{eff}=3/2$ and 2
half-occupied $J_{eff}=1/2$ states. The bands formed by the
$J_{eff}=1/2$ states are much narrower than the width of the whole
$t_{2g}$ manifold in the absence of SOC, and as a result, the system
can be easily drawn to a Mott-insulating phase with even a modest
amount of correlations on the 5d Ir atom.  \cite{zhang2013,kim2009,laguna-marco2010, ju2013}

The most widely studied SOC induced correlated insulator is Sr$_2$IrO$_4$,
which is an antiferromagnetic insulator below 240 K.\cite{moon2008,kim2008}
There are numerous studies that involve strain, and pressure on this
material; and various related compounds are also extensively
studied.\cite{fujiyama2014,gunasekera2014, haskel2012, lee2012,
  liu2014,martins2011,sala2014}
However, despite being the prototypical system, Sr$_2$IrO$_4$ is far
from being the ideal $J_{eff}=1/2$ Mott insulator: the existence of
the insulating state above the Neel temperature is due to short range
order, which is around 100 lattice constants even 20K above the Neel
temperature,\cite{fujiyama2012} hence Sr$_2$IrO$_4$
was termed a \textit{`marginal Mott insulator'}. This marginal nature of the
insulating state was confirmed theoretically, as the first-principles
calculations, which neglect short range order, predict bad metallic
state in the paramagnetic phase.\cite{arita2012}
Also, the crystal structure of Sr$_2$IrO$_4$ is far from
cubic: it has the tetragonal spacegroup I4$_1$/acd. The
tetragonal symmetry breaks the degeneracy of the t$_{2g}$ orbitals, and
thus the $J_{eff}=1/2$ orbitals mix, moving the system away from the
ideal limit where the moments are $SU(2)$ invariant. Since $SU(2)$
symmetric $J_{eff}=1/2$ insulators are proposed to exhibit
superconductivity when doped,\cite{wang2011} it is important
to identify new compounds that are true $J_{eff}=1/2$ Mott insulators
with sizeable gaps.


In this study, we predict a novel class of $J_{eff}=1/2$ Mott
insulator compounds that are both very close to the $SU(2)$ limit and
have large charge gaps in the paramagnetic state. We achieve this by
considering crystal structures that are not commonly studied in the 
context of correlated electron physics. We focus on a group of already
synthesized iridium and rhodium fluoride compounds and use first-principles
calculations at the level of fully charge self-consistent DFT+DMFT
to show the presence of the $J_{eff}=1/2$ insulating state in these compounds. 
We thus expand the search for new $J_{eff}=1/2$ insulators to the family of fluorides, and for the first time show that the $J_{eff}=1/2$ state can exist in a rhodium compound. 

We begin our search for new $J_{eff}=1/2$ insulators by the well known
observation that lower bandwidth favours the Mott insulating
phase. The Sr$_{n+1}$V$_n$O$_{3n+1}$ Ruddlesden-Popper (RP) series nicely
demonstrates this point:\cite{nozaki1991} The $n=\infty$ SrVO$_3$ is a
correlated metal.
In this compound, the oxygen octahedra are corner sharing, and the
number of nearest neighbour transition metal ions is $z=6$.
With decreasing $n$, $z$ decreases monotonically from $z=6$ to $z=4$
for Sr$_2$VO$_4$ ($n=1$).
%
%
This leads to a decrease of the bandwidth as $n$ decreases, and as a
result there is a metal-insulator transition as a function of $n$, and
Sr$_2$VO$_4$ is a Mott insulator.\cite{nozaki1991}
The Sr$_{n+1}$Ir$_n$O$_{3n+1}$ compounds also behave similarly: The
perovskite SrIrO$_3$ ($z=6$) is a correlated metal, the $n=2$
Sr$_3$Ir$_2$O$_7$ ($z=5$) is barely an insulator, and the $n=1$
Sr$_2$IrO$_4$ ($z=4$) is the well-known $J_{eff}=1/2$
insulator.\cite{kim2008, moon2008}

A strategy to obtain a small bandwidth and hence a possible
$J_{eff}=1/2$ Mott insulator in an iridate compound is to look for
crystal structures where the connectivity of anion octahedra is
low. The extreme case is a structure that consists of isolated IrO$_6$
octahedra that are not corner-, edge-, or face- sharing with any other
octahedra. But, to the best of our knowledge, there exists no
structure with isolated MO$_6$ units in transition metal oxides.
However, isolated hexafluoro- transition
metal complexes (MF$_6$) are known to exist and are very common in
fluoride compounds.\cite{wells2012} The Ir ion in
many of these compounds have the d$^5$ electronic configuration, and
hence can lead to the $J_{eff}=1/2$ Mott insulating phase.

\begin{figure}[t]
  \begin{center}
    \includegraphics[width=0.95\hsize]{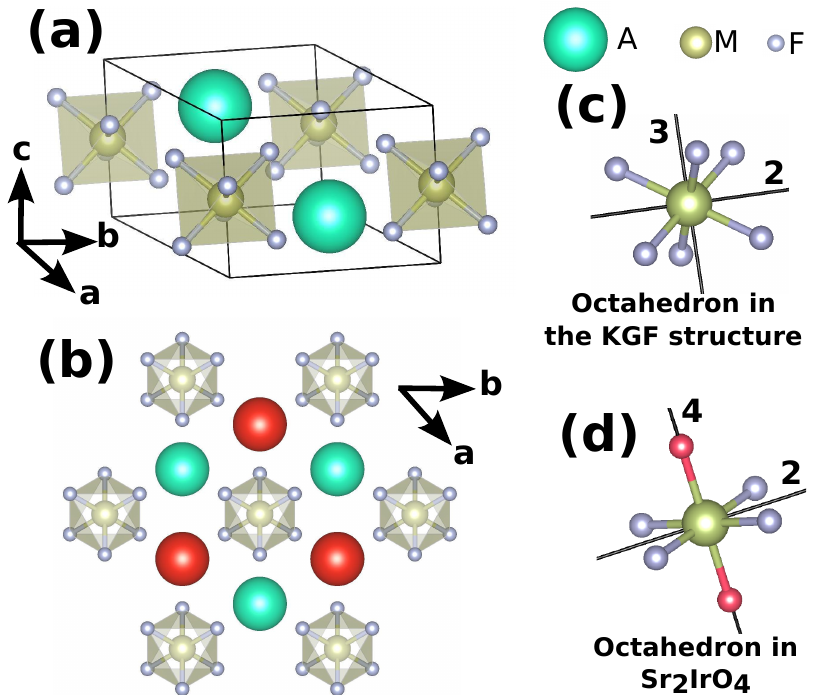}
  \end{center}
    \caption{(a) The K$_2$GeF$_6$ (KGF) crystal structure. (b) The MF$_6$ octahedra are aligned parallel and form triangular layers. The alkali metals are both above and below these layers, shown by green and red. (c) Coordination environment of the transition metals in the KGF and (d) the $n=1$ RP structure. Chemically inequivalent F ions in the RP structure are shown by blue and red.}
    \label{fig:structure}
\end{figure}

As an example of this group of compounds, we consider the alkali metal
hexafluoro-iridates and rhodates with the chemical formula A$_2$MF$_6$
and the so called K$_2$GeF$_6$ (KGF) crystal
structure\cite{babel1967,massa1988} shown in
Fig. \ref{fig:structure}. Here, A is the alkali metal ion and M is the
transition metal ion. Each M ion (in our case either Ir or Rh) is in
the center of an F$_6$ octahedron. The space group is trigonal
P$\bar{3}$m1. While there is no symmetry element that imposes the
octahedra to be regular, all six M-F bondlengths are equal and the
F-M-F angles are close to 90$^\circ$. The site symmetry of the M ion
is $\bar{3}$m, and the threefold degenerate $t_{2g}$ states are split
into 2+1. However, unlike in the RP compound Sr$_2$IrO$_4$, all
ligands are symmetry equivalent (Fig. \ref{fig:structure}c-d), and as
a result an equally distorted octahedron is expected to cause a
smaller splitting of the $t_{2g}$ states in the A$_2$MF$_6$ compounds
than in the RP compounds. 

The M ions form regular triangular layers
(Fig. \ref{fig:structure}b). The octahedra and the local coordinate
axes of all M ions are aligned in a parallel fashion. The out-of-plane 
lattice constant $c$ is
smaller than the in-plane lattice constant $a$, and as a result, the
band structure is of 3-dimensional character. The Ir and Rh cations we
consider have 4+ formal valence and 5 electrons in their $t_{2g}$
orbitals in this structure. Since the MF$_6$ octahedra are isolated in
the sense that there are no F ligands that are coordinated to two
different M ions, the effective hopping between the
M ions is small and hence the $d$ bands at the Fermi level are
expected to be extremely narrow - rendering the system a strong Mott
insulator.

\begin{figure}[t]
  \begin{center}
    \includegraphics[width=1.0\hsize]{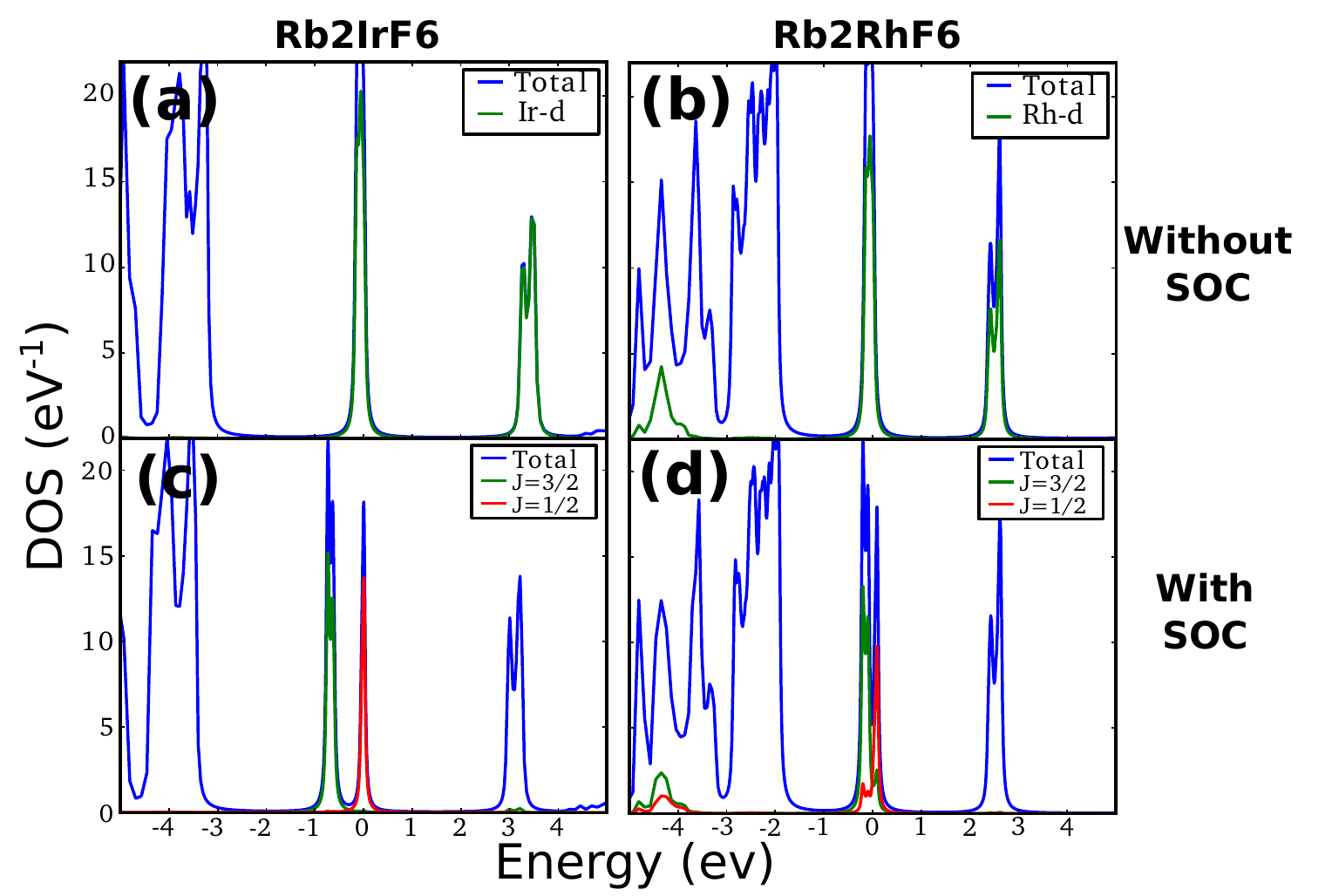}
  \end{center}
    \caption{DOS of Rb$_2$IrF$_6$ and Rb$_2$RhF$_6$ within Density
      Functional Theory, with and without spin-orbit coupling. }
    \label{fig:wienDOS}
\end{figure}

In Fig. \ref{fig:wienDOS}, we show the densities of states (DOS) of
Rb$_2$IrF$_6$ and Rb$_2$RhF$_6$, obtained from Density Functional
Theory in the Generalized Gradient Approximation\cite{PBE} using the
full-potential linear augmented wave formalism as implemented in
WIEN2K\cite{WIEN2K}, and using the experimental crystal
structures.\cite{weise1953, smolentsev2007b} When the SOC is not taken
into account, both compounds have very similar DOS
(Fig. \ref{fig:wienDOS}a-b): There is a narrow ($\sim$400 meV) band
that consists of the transition metal $t_{2g}$ states, which is
partially occupied. The $t_{2g}$-$e_g$ splitting is $\sim$3 eV, and
the $e_g$ states are well above the fermi level. There is no other
state than the $t_{2g}$ states around the Fermi level for a 4-5 eV
interval.

The strong spin orbit coupling due to the heavy Ir ion in
Rb$_2$IrF$_6$ dramatically alters the band structure of this compound
(Fig. \ref{fig:wienDOS}c). The partially filled $t_{2g}$ band near the
Fermi level is split into two bands, a lower lying $J_{eff}=3/2$ band
with 4 electrons, and a half filled $J_{eff}=1/2$ band that crosses
the Fermi level. The latter is extremely narrow ($\sim$100 meV) but
since the Mott physics is beyond DFT, this theory predicts metallic
state. The Rh ion in Rb$_2$RhF$_6$, which is above Ir in
the periodic table, introduces a much weaker SOC than Ir. As a result,
even when SOC is taken into account, the $J_{eff}=3/2$ states are not
energetically separated from the $J_{eff}=1/2$ ones. However, it is
still possible to identify the two overlapping peaks corresponding
to these two groups of states in the DOS.

Both of these compounds have narrow, half filled $J_{eff}=1/2$ bands
near the Fermi level, indicating that a small amount of on-site
correlations can drive them into a Mott-insulating state. While this
state is beyond DFT at the GGA level, it is possible to capture the
Mott insulating phase using Dynamical Mean Field Theory (DMFT).
\footnote{In the DFT+DMFT implementation that we used\cite{haule2010}, a self
energy $\Sigma$ which contains all Feynman diagrams local to the Ir
ion is added to the Kohn-Sham Hamiltonian. The self energy is obtained
by solving the local impurity problem using continuous time quantum
Monte Carlo\cite{CTQMC1,CTQMC2} and full charge self-consistency is
obtained by repeating DFT and DMFT steps.}
DFT+DMFT has been successfully applied to reproduce the properties of
various Mott insulators and it has been recently used to study the
$J_{eff}=1/2$ insulating phase in Sr$_2$IrO$_4$.\cite{zhang2013}
As a result, it is the natural method of choice to study the possibly
Mott insulating electronic structure of the hexafluoro-iridates and
-rhodates. We chose the same on-site Coulomb repulsion in these
compounds as estimated for iridates in Ref.~\onlinecite{zhang2013}, i.e.,
$U=4.5\,$eV and $J=0.8\,$eV. We note that these values are the lower
bound for more localized fluorides, hence we are possibly
underestimating the size of the Mott gap.

\begin{figure}[t]
  \begin{center}
    \includegraphics[width=0.9\hsize]{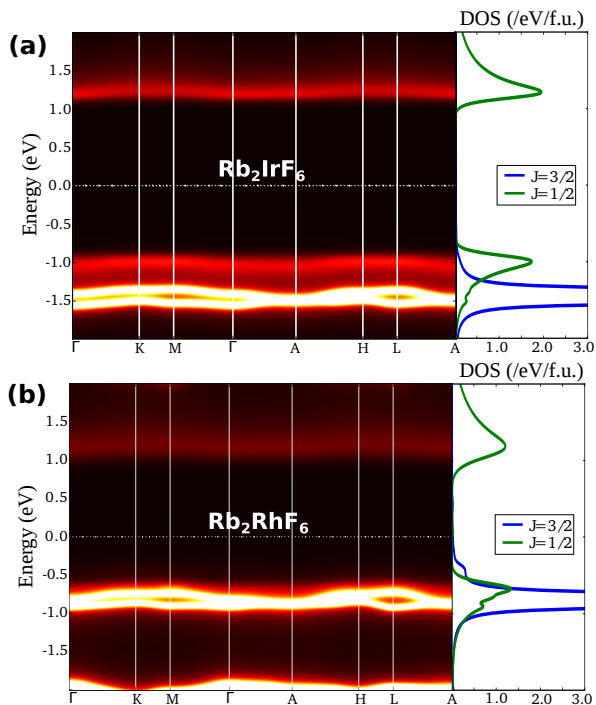}
  \end{center}
    \caption{The spectral function A(k,$\omega$) and DOS of (a) Rb$_2$IrF$_6$, and (b) Rb$_2$RhF$_6$ (bottom).}
    \label{fig:Akw_Rb}
\end{figure}

In Fig. \ref{fig:Akw_Rb}, we present the result of our DMFT
calculations: The spectral functions A(k,$\omega$) of Rb$_2$IrF$_6$
and Rb$_2$RhF$_6$ from DFT+DMFT.\footnote{In our DMFT calculations, 
the temperature is taken to be 0.01 eV. While
  in the beginning of the calculations the symmetry of the self energy
  is broken and magnetic ordering is allowed, it became spin-symmetric
  after a few iterations, indicating that the compounds do not have
  any strong tendency towards magnetic ordering at this temperature.}
Both compounds are Mott insulators, with wide gaps close to
$\sim$2 eV. In Rb$_2$IrF$_6$, the upper and lower Hubbard bands are
clearly separated and have $J_{eff}=1/2$ character, indicating that
Rb$_2$IrF$_6$ is a $J_{eff}=1/2$ Mott insulator. In Rb$_2$RhF$_6$, the
lower Hubbard band overlaps with the fully occupied, uncorrelated
$J_{eff}=3/2$ bands and so cannot be clearly seen in the A(k,$\omega$)
plot. However, the upper Hubbard band has a clear $J_{eff}=1/2$
character and therefore this compound is a $J_{eff}=1/2$ insulator as
well.
To the best of our knowledge, this is the first report of a
$J_{eff}=1/2$ insulator in a compound that does not contain
Iridium. Furthermore, both of these compounds have the largest gaps
ever reported for a $J_{eff}=1/2$ insulator. Both the large gaps, and
the possibility of the $J_{eff}=1/2$ state in a rhodate compound are
thanks to the non-connectivity of the MF$_6$ octahedra in the KGF
structure, and the resulting very narrow $J_{eff}=1/2$ bands.

In passing, we note that replacing Ir with Rh in Sr$_2$IrO$_4$ leads to a metallic phase both because of the much weaker SOC\cite{qi2012} but also possibly because of the slightly larger electronegativity of the Rh ion. In the KGF fluorides, the charge gap is large, which makes the electronegativity difference negligable, and also the SOC is not necessary for the Mott insulating phase (it is essential only for the J$_{eff}$=1/2 character). As a result, even the rhodates in this structure are J$_{eff}$=1/2 Mott insulators. (See the supplamental material.)  

Encouraged by the success of our strategy to look for $J_{eff}=1/2$
Mott insulators in this class of coumpounds, we also performed
DFT+DMFT calculations in three other compounds with the same crystal
structure, Cs$_2$IrF$_6$, K$_2$IrF$_6$, and K$_2$RhF$_6$. While these
compounds have significantly different lattice constants due to the
different alkali metals they contain, we find all of them to be
$J_{eff}=1/2$ insulators with large gaps as well.
All of these compounds were synthesized and their crystal structures
were studied
before\cite{weise1953,hepworth1958,fitz2002,smolentsev2007a,smolentsev2007b,
  babel1967,massa1988} but there is very little information on their
magnetic properties or conductivities. Our predictions call for more
experiments to characterize these materials better.
We predict a fluctuating magnetic moment of 1.6 $\mu_B$ in both
Rb$_2$IrF$_6$ and Rb$_2$RhF$_6$, which is smaller than the value
expected for an ideal spin-1/2 Mott insulator (1.73 $\mu_b$) because
of the charge fluctuations (there is a $>10 \%$ probability that there
are 6 electrons in the $t_{2g}$ orbitals). These values are $\sim 12\%$ larger than
what is measured in Cs$_2$IrF$_6$ in Ref. \onlinecite{hepworth1958},
but $\sim 8\%$ smaller than the value measured in the Rh compounds in
Ref. \onlinecite{weise1953} at room temperature.
%

The ideal $J_{eff}=1/2$ state is $SU(2)$ invariant, and so it has no
magnetic anisotropy. However, systems such as Sr$_2$IrO$_4$ are
observed not to be exactly at this limit due to deviations of the wave
function from the ideal $J_{eff}=1/2$.\cite{fujiyama2014} The reason
is that Sr$_2$IrO$_4$ lacks cubic symmetry: It has the space group
I4$_1$/acd, which is tetragonal, and hence the three $t_{2g}$ orbitals
of the Ir ion are split into a degenerate doublet and a singlet.
The deviation from ideal $J_{eff}=1/2$ state is small but not
negligible, and it depends strongly on biaxial strain and
pressure.\cite{zhang2013, serrao2013, haskel2012} In the KGF
structure, the space group is trigonal, and the $t_{2g}$ irreducible
representation is split into two, a singlet $A_{1g}$ and a doublet
$E_g$, similar to Sr$_2$IrO$_4$. This also introduces a deviation from the $J_{eff}=1/2$
state and a resultant magnetic anisotropy.

In order to see how much the wavefunction is different from the ideal
$J_{eff}=1/2$ state, we study the hybridization function
$\Delta(\omega)$ used in the DMFT calculation.\cite{haule2010} It is
given by
\begin{equation}
\frac{1}{\omega - \Delta 
(\omega)-\Sigma(\omega)}=\sum_{\vec{k}}\hat{P}_{\vec{k}}\frac{1}{\omega+\mu-\epsilon_{\vec{k}}-\hat{P}_{\vec{k}}^{-1}\Sigma(\omega)}
\end{equation}
where $\Sigma(\omega)$ is the DMFT self energy, $\epsilon_{\vec{k}}$
are the DFT Kohn-Sham eigenvalues, and $\hat{P}$ and $\hat{P}^{-1}$
are the projector and the embedder on the transition metal site. In
the high frequency limit $\omega\rightarrow\infty$, the eigenvalues of
the $\Delta$ matrix give the atomic energy levels (including both the
crystal field and the spin-orbit coupling) and it is related to the
single ion anisotropy. In the $\omega \rightarrow 0$ limit, it is
related to the low energy electronic excitations. The two eigenvectors
of $\Delta$ with the largest eigenvalues are the $J_{eff}=1/2$-like states $|\psi_{+1/2}\rangle$ and $|\psi_{-1/2}\rangle$. The inner products of these with the ideal $J_{eff}=1/2$ states 
$|J_{1/2,\mp 1/2}\rangle$
can be used as a measure of how close the system to the $SU(2)$ limit is. However, this product is second order in the mixing, and a better measure is the coefficients in the expansions of $|\psi{_{\mp1/2}}\rangle$. This measure is used in Ref. [\onlinecite{zhang2013}] to study the effect of tetragonal symmetry breaking in Sr$_2$IrO$_4$. 
%
%
Under a trigonal perturbation, the $t_{2g}$ orbitals are split into a singlet and a doublet as\cite{yang2010}
%
%
$|a_1\rangle=\frac{1}{\sqrt{3}}\left(|d_{xy}\rangle +|d_{yz}\rangle +|d_{xz}\rangle\right)$, 
$|e_+\rangle=\frac{1}{\sqrt{3}}\left(|d_{xy}\rangle +\alpha|d_{yz}\rangle +\alpha^2|d_{xz}\rangle\right)$, and $|e_-\rangle=\frac{1}{\sqrt{3}}\left(|d_{xy}\rangle +\alpha^2|d_{yz}\rangle +\alpha|d_{xz}\rangle\right)$, 
where $\alpha=e^{i2\pi/3}$. A generalization of the $J_{eff}=1/2$ states that takes into account this splitting is 
\begin{widetext}
\begin{equation}
|\psi_{+1/2}\rangle=\frac{\sqrt{3-2\gamma^2}}{3}\Big(-|a_1\downarrow\rangle+(1-i)|a_1\uparrow\rangle\Big)
+\frac{\gamma}{3}\Big(
\big[-|e_+\downarrow\rangle+(\alpha^2-i\alpha)|e_+\uparrow\rangle\big]
+\big[-|e_-\downarrow\rangle+(\alpha-i\alpha^2)|e_-\uparrow\rangle\big]
\Big)
\end{equation}
\begin{equation}
|\psi_{-1/2}\rangle=\frac{\sqrt{3-2\gamma^2}}{3}\Big(-|a_1\uparrow\rangle+(1+i)|a_1\downarrow\rangle\Big)
+\frac{\gamma}{3}\Big(
\big[|e_+\uparrow\rangle+(\alpha^2+i\alpha)|e_+\downarrow\rangle\big]
+\big[|e_-\uparrow\rangle+(\alpha+i\alpha^2)|e_-\downarrow\rangle\big]
\Big)
\end{equation}
\end{widetext}
Here, $\gamma$ quantifies the deviation from the ideal limit, and
$\gamma=1$ gives $|\psi_{\mp 1/2}\rangle=|J_{1/2,\mp
  1/2}\rangle$. 
A large $|1-\gamma|$ indicates strong deviation from the Heisenberg regime, and leads to large magnon gaps, even larger than the spin wave bandwidth in Sr$_3$Ir$_2$O$_7$.\cite{kim2012,zhang2013}
%
Since hybridization is frequency dependent, so is
$\gamma$.
In Rb$_2$IrF$_6$, the low frequency $\gamma_0=0.987$ and the high
frequency $\gamma_\infty=0.992$. Compared to Sr$_2$IrO$_4$,\cite{zhang2013} which
has $\gamma_0=1.03$ and $\gamma_\infty=1.02$; the
electronic state in Rb$_2$IrF$_6$ is much more isotropic, and closer
to the ideal $SU(2)$ limit. The rhodate compound Rb$_2$RhF$_6$, which
has weaker SOC, shows a more significant deviation from the ideal
limit: It has $\gamma_0=1.020$ and $\gamma_\infty=0.935$.

This very isotropic behaviour despite the noncubic spacegroup of
Rb$_2$IrF$_6$ can be better understood considering the local
coordination geometry of the transition metal ion. The site symmetry
of Ir is $\bar{3}m$. The elements of the point group include various
rotations, such as a threefold rotation around [001] and a twofold
rotation around [100] (Fig. \ref{fig:structure}c). As a result, all
six F-ligands around a M ion are symmetry equivalent: They are
chemically identical, and their F-M bond lengths are the same. The
deviation from the ideal cubic symmetry on the M site is only due to
the presence of further neighbours that reduce the symmetry, and the
deviation of the F-octahedra from a regular octahedron. This latter
effect is quite small (the largest F-M-F bond angle variance in the
compounds we consider is less than 6 degrees), and as a result, the Ir
ion is in an almost cubic environment. In the RP family of iridate
compounds, the site symmetry of the Ir ion can be as high as
$4/mmm$. However, despite a 4-fold rotation and various 2-fold
rotation axes that pass through the Ir ion
(Fig. \ref{fig:structure}d), there is no 3-fold rotation in the point
group, and there are two \textit{chemically} distinct ligands around
each Ir ion. The apical oxygens, shown by red in
Fig. \ref{fig:structure}d, are bonded to only one Ir ion, whereas the
other oxygens are bonded to two Ir each. This necessarily results in
very a noncubic \textit{local} environment of the Ir ion, which leads
to deviations from the ideal $J_{eff}=1/2$ state even when the Ir-O
bondlengths are artificially set to be equal.

In conclusion, we identified a new class of $J_{eff}=1/2$ Mott
insulators, which includes the first two examples of such compounds
without iridium. These materials are wide gap Mott insulators, with no
visible tendency towards magnetic ordering, and some of them are also
closer to the isotropic $SU(2)$ limit than the well studied
Sr$_2$IrO$_4$. This work extends the search for new materials which
display an interplay of correlations with spin-orbit coupling to
flouride compounds.
We posit that the $J_{eff}=1/2$ Mott insulating phase is very common
in transition metal fluorides with isolated Ir$^{4+}$F$_6$ and
Rh$^{4+}$F$_6$ complexes. Studying other structure types that satisfy
this property would lead not only to the discovery of new
$J_{eff}=1/2$ Mott insulators but also many other strongly correlated
complex fluorides with interesting physical properties.

\textit{Note:} While this manuscript was under review, we became aware of a study on RuCl$_3$ which also reports a relativistic Mott insulating phase in a 4d transition metal halide.\cite{plumb2014}

We acknowledge fruitful discussions with N.A. Benedek and O. Erten. T. B. was supported by the Rutgers Center for Materials Theory, and K. H. was supported by NSF DMR-1405303 and NSF DMREF-1233349.

\pagebreak
\widetext
\setcounter{equation}{0}
\setcounter{figure}{0}
\setcounter{table}{0}
\setcounter{page}{1}
\makeatletter
\renewcommand{\theequation}{S\arabic{equation}}
\renewcommand{\thefigure}{S\arabic{figure}}
\renewcommand{\bibnumfmt}[1]{[S#1]}
\renewcommand{\citenumfont}[1]{S#1}

\begin{center}
\textbf{Supplemental Information for\\``The $J_{eff}=1/2$ Mott Insulating State in Rh and Ir Fluorides"}
\end{center}

\section{Details of the DMFT Implementation}
The action of the auxiliary impurity problem that we minize in our calculations is 
\begin{multline}\label{equ:action}
\mathcal{S}=\int_0^\beta d\tau\psi^\dagger_{L\sigma}(\tau)\frac{\partial}{\partial\tau} \psi_{L\sigma}(\tau) +\int_0^\beta d\tau \int_0^\beta d\tau'\psi^\dagger(\tau')\Delta_{L_1\sigma,L_2\sigma'}\psi(\tau)\\
+\frac{1}{2}\int_0^\beta d\tau \sum_{L_1,L_2,L_3,L_4,\sigma,\sigma'} U_{L_1,L_2,L_3,L_4}\psi^\dagger_{L_1 \sigma}(\tau) \psi^\dagger_{L_2 \sigma'}(\tau) \psi_{L_3 \sigma'}(\tau) \psi_{L_4 \sigma}(\tau) 
\end{multline}
where $\tau$ is the imaginary time, $\psi$ is the annihilation operator for the impurity electrons, $L$ and $\sigma$ are the orbital and spin indices, and $\Delta$ is the impurity hybridization function. The on site electron-electron Coulomb interaction between the d electrons is represented by the Slater form:
\begin{equation}
\hat{U}=\frac{1}{2}\sum_{L_1,L_2,L_3,L_4,\sigma,\sigma'} U_{L_1,L_2,L_3,L_4}c^\dagger_{L_1 \sigma} c^\dagger_{L_2 \sigma'} c_{L_3 \sigma'} c_{L_4 \sigma}
\end{equation}
and
\begin{equation}
U_{L_1,L_2,L_3,L_4}=\sum_k \frac{4 \pi}{2k+1}\langle Y_{L_1}|Y_{km}|Y_{L_4}\rangle\langle Y_{L_2}|Y^*_{km}|Y_{L_3}\rangle F^k_{l_1,l_2,l_3,l_4}
\end{equation}
Here, $Y$ are the radial part of the spherical harmonics, the index $L$ denotes $L\rightarrow \{l,m\}$, and $F^k$ are the Slater integrals. The explicit form of the frequency dependent hybridization $\Delta(i\omega)$ can be written in terms of the projector $P$, the Kohn-Sham Hamiltonian (without the spin-orbit coupling) $\mathcal{H}^{DFT}$, the spin-orbit coupling Hamiltonian $\mathcal{H}^{SOC}$, the self-energy $\Sigma(i\omega)$, the local Green's function $G_{loc}(i\omega)$ and the double counting energy $V_{DC}$ as\cite{Shaule2010}
\begin{equation}
\Delta_{L\sigma,L'\sigma}(i\omega)=\sum_{\vec{k}, i, j} P_{\vec{k}}(L\sigma, L'\sigma', ij)
(\mathcal{H}^{DFT}+\mathcal{H}^{SOC})+i\omega-\Sigma(i\omega)-G^{-1}_{loc}(i\omega)-V_{dc}
\end{equation}
The small latin indices $i, j$ enumerate the Kohn-Sham bands and include the spin as well. $\vec{k}$ is the crystal momentum. The form of the spin-orbit coupling Hamiltonian is $\mathcal{H}^{SOC}\sim \lambda \vec{L}\cdot\vec{S}$. Forms of different projectors ($P$) and double counting ($V_{DC}$) have been previously discussed in, for example, [\onlinecite{Shaule2010}] and [\onlinecite{Shaule2014}]. $\Sigma$, $G_{loc}$, $\mathcal{H}^{DFT}$, and $\mathcal{H}^{SOC}$ are determined self consistently by extremizing the action (\ref{equ:action}), hence $\Sigma$ and $G_{loc}$ obey the DMFT self consistency condition and $\mathcal{H}^{DFT}$, and $\mathcal{H}^{SOC}$ are determined from the self consistent electronic charge. 

The impurity model is solved with the continous time quantum monte carlo (CTQMC) solver. In order to reduce the sign problem, we employ orbital and spin rotations ($\mathcal{U}$) that cast the hybridizaton $\Delta$ in a diagonal form in the relevant low energy limit 
\begin{equation}
\Delta(i\omega) = \mathcal{U}^\dagger \Delta_{diag}(i\omega) \mathcal{U}+\delta \Delta(i\omega)
\end{equation}
The hybridization is strongly frequency dependent, however, its eigenvectors do not change much with frequency. In other words, it is possible to obtain an \textit{almost diagonal hybridization} by using a \textit{frequency independent} rotation $\mathcal{U}$. We ignore the off diagonal $\delta\Delta(i\omega)$ and as a result our CTQMC impurity solver suffers very little sign problem, even in the presence of spin-orbit coupling and low crystal symmetry.

\section{The Effect of Spin-Orbit Coupling on the Insulating Behaviour}

Unlike  Sr$_2$IrO$_4$, where spin-orbit coupling (SOC) is necessary for the insulating behaviour, the compounds with the KGF structure that we consider can be Mott insulators even in the absence of SOC, because they have narrow bands in DFT even without SOC. In order to verify this point, we performed DFT+DMFT calculations without SOC on Rb$_2$RhF$_6$ and Rb$_2$IrF$_6$. The DOS from these calculations, along with the DOS with SOC for comparison, are presented in Figure 1. As expected from a crystal structure with unconnected polyhedra, both of these compounds are Mott insulators with wide charge gaps. Rb$_2$IrF$_6$ has a smaller gap when SOC is not taken into account, but the width of the gap of Rb$_2$RhF$_6$ does not change much, in line with the small energy scale of the SOC in this compound. 

As a result, we conclude that the SOC in the TM-fluorides with the KGF structure is not responsible of the insulating behaviour, but it only changes the character of the insulating state. In the absence of SOC, these compounds would be spin-1/2 Mott insulators, whereas, due to SOC, they become J$_{eff}$=1/2 Mott insulators. This difference is clear considering the character of the low energy degrees of freedom, and would be visible in magnetic properties, such as the stength of the magnetocrystalline anisotropy or the magnon dispersion. 

\begin{figure}[t]
  \begin{center}
    \includegraphics[width=0.9\hsize]{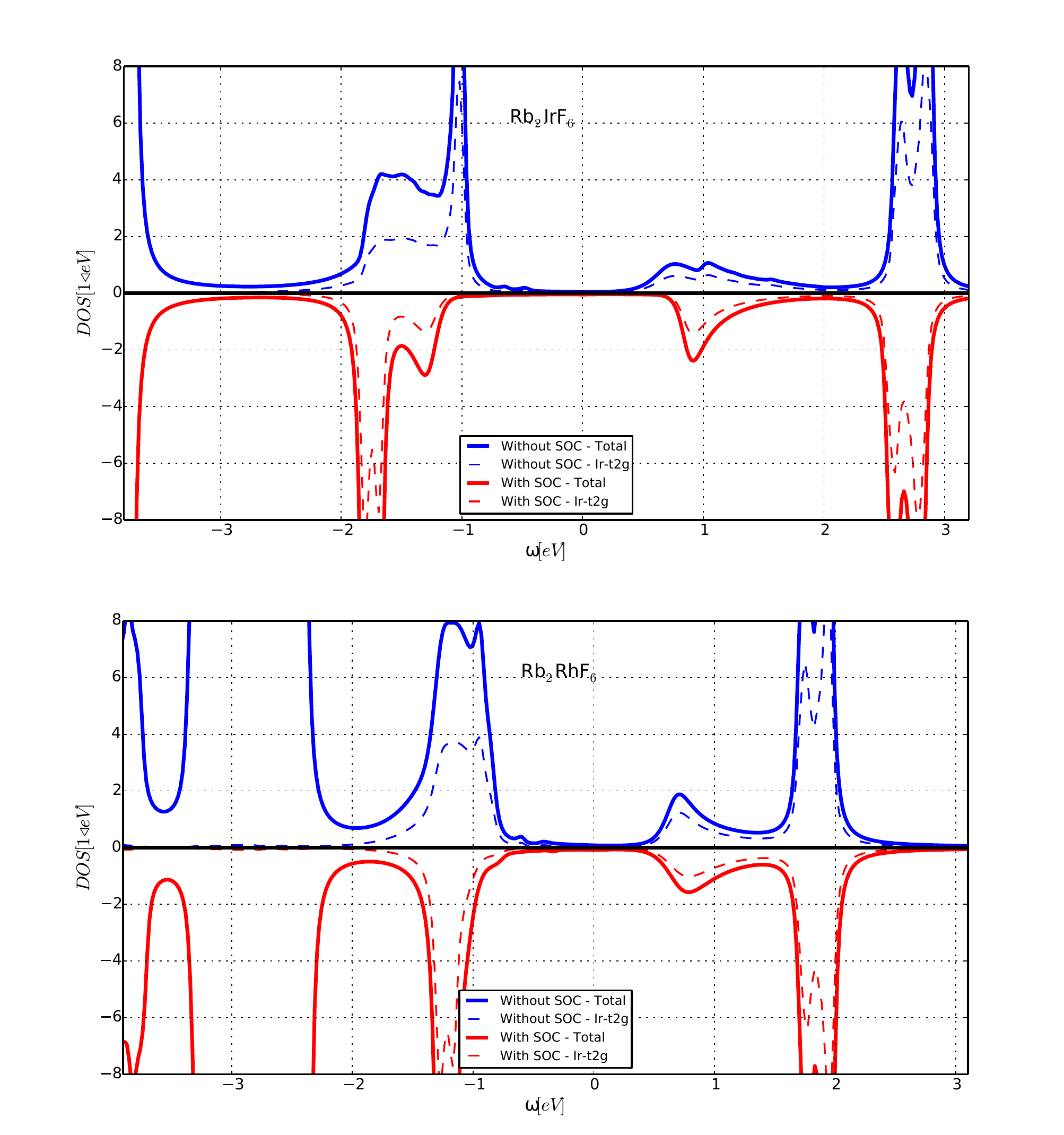}
  \end{center}
    \caption{The densities of states for Rb$_2$IrF$_6$ and Rb$_2$RhF$_6$ with and without spin orbit coupling. }
    \label{fig:Akw_Rb}
\end{figure}

\section{Magnetic Moments' Dependence on $\gamma$ and the Deviation From Heisenberg Behaviour}
In the main text, we generalized the $|J_{1/2,\mp 1/2}\rangle$ states as 
\begin{equation}
|\psi_{+1/2}\rangle=\frac{\sqrt{3-2\gamma^2}}{3}\Big(-|a_1\downarrow\rangle+(1-i)|a_1\uparrow\rangle\Big)
+\frac{\gamma}{3}\Big(
\big[-|e_+\downarrow\rangle+(\alpha^2-i\alpha)|e_+\uparrow\rangle\big]
+\big[-|e_-\downarrow\rangle+(\alpha-i\alpha^2)|e_-\uparrow\rangle\big]
\Big)
\end{equation}
\begin{equation}
|\psi_{-1/2}\rangle=\frac{\sqrt{3-2\gamma^2}}{3}\Big(-|a_1\uparrow\rangle+(1+i)|a_1\downarrow\rangle\Big)
+\frac{\gamma}{3}\Big(
\big[|e_+\uparrow\rangle+(\alpha^2+i\alpha)|e_+\downarrow\rangle\big]
+\big[|e_-\uparrow\rangle+(\alpha+i\alpha^2)|e_-\downarrow\rangle\big]
\Big)
\end{equation}
where 
\begin{equation}
|a_1\rangle=\frac{1}{\sqrt{3}}\left(|d_{xy}\rangle +|d_{yz}\rangle +|d_{xz}\rangle\right)
\end{equation}
\begin{equation}
|e_+\rangle=\frac{1}{\sqrt{3}}\left(|d_{xy}\rangle +\alpha|d_{yz}\rangle +\alpha^2|d_{xz}\rangle\right)
\end{equation}
\begin{equation}
|e_-\rangle=\frac{1}{\sqrt{3}}\left(|d_{xy}\rangle +\alpha^2|d_{yz}\rangle +\alpha|d_{xz}\rangle\right)
\end{equation}
and $\alpha=e^{i2\pi/3}$. These states are the eigenstates of the SOC Hamiltonian under a trigonal distortion, which breaks the degeneracy of $|a_1\rangle$ with $|e_\mp\rangle$. 
The ideal J$_{eff}$=1/2 states $|J_{1/2,\mp 1/2}\rangle$ are SU(2) invariant: the ratio of orbital to spin angular momenta is $\langle \mu_L \rangle/\langle\mu_S\rangle=2$ and is independent of direction. However, the generalized states $|\psi_{\mp 1/2}\rangle$ do not satisfy these conditions. For example, for $z$ being the axis of trigonal distortion, and $x=\gamma-1\ll 1$, one gets $\langle \mu_L^z \rangle/\langle\mu_S^z\rangle=2-2x^2$. In this respect, $\gamma$ is a natural parameter to quantify the deviation of the system from an ideal J$_{eff}$=1/2 insulator. 

The fact that the magnetic moment is anisotropic, and that the system is not in a Heisenberg regime, has important consequences in the magnetic excitation spectrum. For example, Sr$_3$Ir$_2$O$_7$, which has a $|1-\gamma|$ that is much larger than Sr$_2$IrO$_4$, has a large (92 meV) magnon gap that is even wider than the magnon bandwidth in this compound.\cite{Szhang2013,Skim2012}

\end{document}